\documentclass[12pt,a4paper,dvips]{article}
\usepackage{a4p}
\usepackage{cite,mcite,epsfig}
\usepackage{graphicx}
\usepackage{physics}
\usepackage{l3_title}
\usepackage{ifthen}
\journalname{Phys. Lett. B}
\date{May 16, 2002}
\preprint{2002-035}
\def\ifmath#1{\relax\ifmmode #1\else $#1$\fi}%
\def\rd{\ifmath{{\mathrm{d}}}}

\def\re{\ifmath{{\mathrm{e}}}}

\def\ev{\ifmath{{\mathrm{ev}}}}
\def\triplets{\ifmath{{\mathrm{triplets}}}}
\def\genuine{\ifmath{{\mathrm{genuine}}}}

\def\mix{\ifmath{{\mathrm{mix}}}}
\def\MC{\ifmath{{\mathrm{MC}}}}
\def\noBE{\ifmath{{\mathrm{noBE}}}}

\def\det{\ifmath{{\mathrm{det}}}}

\def\exp{\ifmath{{\mathrm{exp}}}}
\def\Re{\ifmath{{\mathrm{Re}}}}

\def\where{\ifmath{{\mathrm{where}}}}

\newlength{\capindent}
\newlength{\capwidth}
\newlength{\figwidth}
\newcommand{\icaption}[2][!*!,!]{\hspace*{\capindent}%
  \begin{minipage}{\capwidth}
    \ifthenelse{\equal{#1}{!*!,!}}%
      {\caption{#2}}%
      {\caption[#1]{#2}}
  \end{minipage}}

\begin{document}
\begin{titlepage}
\title{Measurement of\\ Genuine Three-Particle Bose-Einstein Correlations\\ in Hadronic Z Decay}
\author{L3 Collaboration}
%
%
\begin{abstract}
We measure three-particle Bose-Einstein correlations 
in hadronic Z decay with the L3 detector at LEP. 
Genuine three-particle Bose-Einstein correlations
are observed. By comparing two- and three-particle
correlations we find that the data are consistent 
with fully incoherent pion production.
\end{abstract}
%
%
\submitted
\end{titlepage}
%
%

\section*{Introduction}
So far, no theory exists which can describe the non-perturbative process of hadron production
in general and Bose-Einstein (BE) effects in particular. The latter are expected from general
spin statistics considerations. To help 
understand these phenomena, studies of identical-boson correlations in $\re^{+}\re^{-}$
collisions at LEP have been performed in terms of the absolute four-momentum difference $Q$~\cite{ee1}, as
well as in two- and three-dimensional distributions in components of $Q$~\cite{elong,elongopdel}.

It has long been realized that the shape and size in space-time of a source of pions can be determined,
as a consequence of the interference of identical bosons, from the shape and size of the correlation
function of two identical pions in energy-momentum space~\cite{basic12}.
Additional information can be derived from higher-order correlations. Furthermore, such correlations
constitute an important theoretical issue for the understanding of Bose-Einstein correlations (BEC)~\cite{biyajima}.

In this Letter three-particle correlations are analysed. These correlations are sensitive to
asymmetries in the particle production mechanism~\cite{heiselbheiselb2,zhang} which 
cannot be studied by two-particle correlations. 
In addition, the combination of two- and three-particle correlation analyses gives access to
the degree of coherence of pion production~\cite{lorstad,andreev}, which is
very difficult to investigate from two-particle correlations alone 
due to the effect of long-lived resonances on the correlation function.
The DELPHI~\cite{delphicol} and OPAL~\cite{opalcol} collaborations have both
studied three-particle correlations but did not investigate the degree of coherence.

\section*{The Data and Monte Carlo}\label{chargedhadsel}
The data used in this analysis were collected by the L3 detector~\cite{l3} in 1994 at a centre-of-mass
energy of 91.2\,\GeV\ and correspond to a total integrated luminosity of 48.1 pb$^{-1}$.
The Monte Carlo (\MC) event generators JETSET~\cite{jetset} and HERWIG~\cite{herwig} are
used to simulate the signal process. Within JETSET, BEC are simulated using 
the BE$_{0}$ algorithm~\cite{jetset0,jetset32}~\footnote{
The Bose-Einstein simulation is done by the subroutine LUBOEI, with the values PARJ(92)=1.5
and PARJ(93)=0.33\,\GeV.}.
The generated events are passed through the L3 detector
simulation program, which is based on the {\sc GEANT}\cite{geant3} and {\sc GHEISHA}\cite{gheisha}
programs, reconstructed and subjected to the same selection criteria as the data.

The event selection is identical to that presented in Reference~\citen{elong}, resulting in about one million
hadronic Z decay events, with an average track multiplicity of about~12.
Two additional cuts are performed
in order to reduce the dependence of the detector correction on the \MC\ model used:
tracks with measured momentum greater than 1\,\GeV\ are rejected, as are pairs of like-sign tracks with
opening angle below~3$^{\circ}$. This results in an average track multiplicity of about~7.
For the computation of three-particle correlations, each
possible triplet of like-sign tracks is used to compute the variable \mbox{$Q_{3}\equiv
\sqrt{Q_{12}^2+Q_{23}^2+Q_{31}^2}$, where $Q_{ij}\equiv\sqrt{-(p_{i}-p_{j})^2}$} is
the absolute four-momentum difference between particles $i$ and $j$. Since $Q_{ij}$,
and thus $Q_3$, depends both on the energy of the particles and on the angle between them, small $Q_{ij}$
can be due to small angles or low energies. In a \MC\ generator with BE effects, 
the fraction of pairs
at small $Q_{ij}$ with small angle is larger than in one without. Consequently, the estimated 
detection efficiency depends on the \MC\  model used.
The momentum and opening angle cuts reduce this model dependence.
After selection, the average triplet multiplicity is about~6.
In the region of interest, $Q_{3}< 1\,\GeV$, the loss of triplets by the 
momentum and opening angle cut is about~40\%.

The momentum cut improves the resolution of $Q_3$ by 
a factor three with respect to that for the full momentum spectrum. Using \MC\  events, its average is 
estimated to be 26\,\MeV\ for triplets of tracks with $Q_3 < 0.8\,\GeV$.
We choose a bin size of 40\,\MeV, somewhat larger than this resolution.

\begin{figure}
\begin{center}
\epsfig{figure=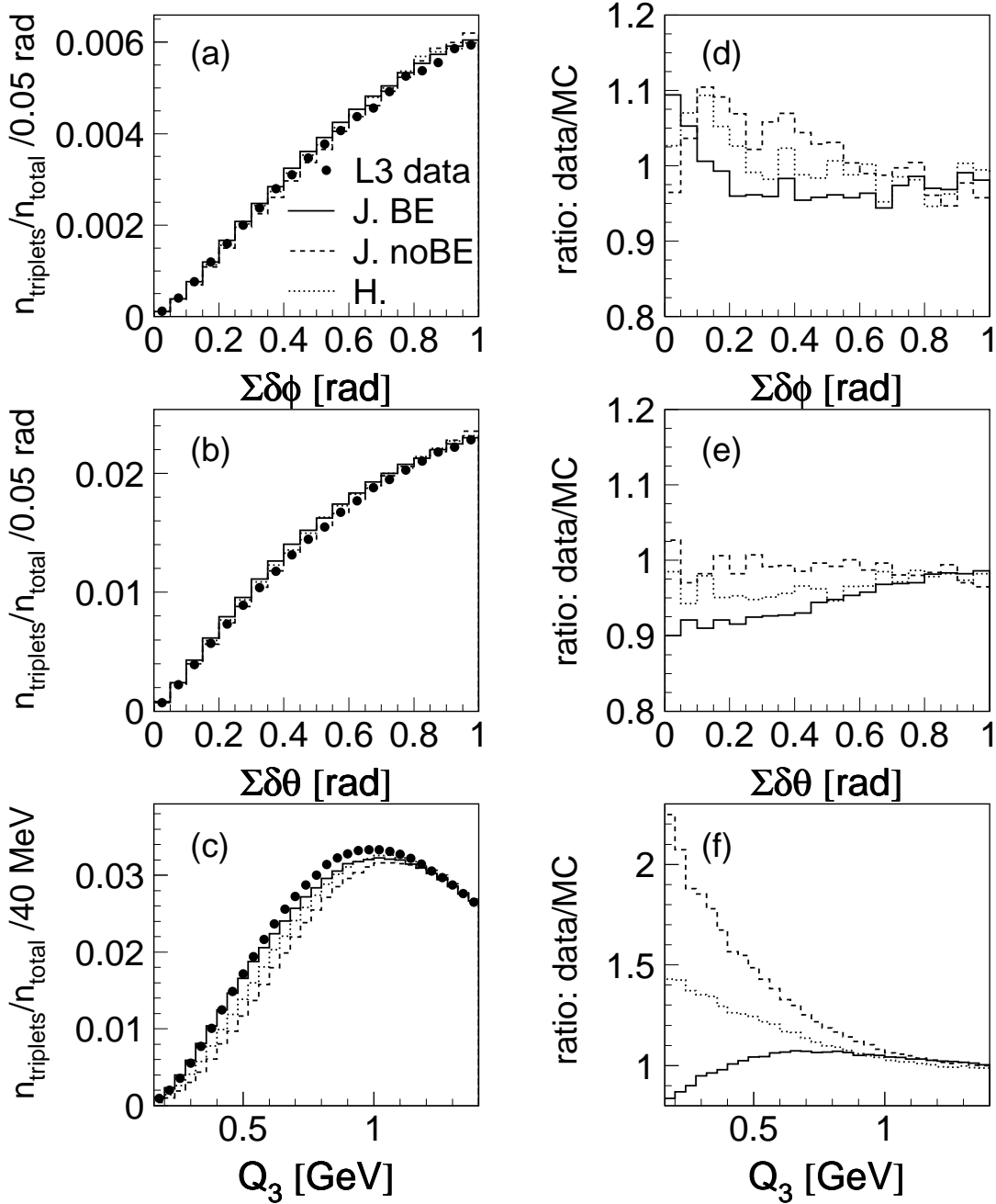, width=.99\linewidth}
\caption{Normalized distributions of the sum of the difference in (a) azimuthal and (b) polar angle of pairs
of tracks in a triplet, $\sum\delta\phi$ and $\sum\delta\theta$, and of (c) $Q_3$.
Data, JETSET, with and without BE effects, and HERWIG are displayed.
The ratios between the data and \MC\  distributions are shown in (d), (e) and (f).}
\label{fig1b}
\end{center}
\end{figure}

In Figure~\ref{fig1b}, the data are compared to JETSET (with and without BE effects) and HERWIG
(not having a BE option) at the detector
level, after performing all the cuts mentioned above, in the three-particle distributions $\sum\delta\phi$,
$\sum\delta\theta$, and $Q_3$. The sums run over the three pairs of like-sign tracks in the triplet and
$\delta\phi$ and $\delta\theta$ are the absolute differences in azimuthal and polar angle between two tracks, respectively.
Within 10\%, the angular distributions of the \MC\  models agree with those of the data. 
None of the models describes the $Q_3$ distribution: 
JETSET with BE effects overestimates the data by approximately 20\% at low $Q_3$, 
even though we found good agreement for
$\delta\phi$,
$\delta\theta$, and $Q$~\cite{elong}.
JETSET without BE effects and HERWIG grossly underestimate the data at low $Q_3$. The statistics
for $Q_3< 160\,\MeV$ are so poor, that this region is rejected from the analysis.

\section*{The Analysis}
The three-particle number density $\rho_{3}(p_1,p_2,p_3)$ of
particles with four-momenta $p_1, p_2$ and $p_3$ can be described in terms
of single-particle, two-particle and genuine three-particle densities as
\begin{equation}
\rho_{3}(p_1,p_2,p_3)=\rho_{1}(p_1)\rho_{1}(p_2)\rho_1(p_3)+\sum_{(3)}\left\{ \rho_{1}(p_1)\left[
\rho_2(p_2,p_3)-\rho_1(p_2)\rho_1(p_3)\right] \right\} + C_{3}(p_1,p_2,p_3)\ \ ,
\label{eqr3}
\end{equation}
where the sum is over the three possible permutations and $C_{3}$ is the third-order cumulant,
which measures the genuine three-particle correlations. 
The $\rho_1\rho_2$ terms contain all the two-particle correlations. 
In order to focus on the correlation due to BE interference, 
we replace products of single-particle densities by the corresponding two- or three-particle
density, $\rho_0$, which would occur in the absence of BEC, 
and define the correlation functions
\begin{equation}
R_{2} (p_1 ,p_2      )\equiv \frac{\rho_{2}(p_1,p_2)}{\rho_{0}(p_1,p_2)} \ \ , \qquad\qquad
R_{3} (p_1 ,p_2, p_3 )\equiv \frac{\rho_{3}(p_1,p_2,p_3)}{\rho_{0}(p_1,p_2,p_3)} \ \ \ .
\label{threecorr}
\end{equation}
Assuming the absence of two-particle correlations, \ie, $\rho_2(p_1,p_2)=\rho_1(p_1)\rho_1(p_2)$, results in
\begin{equation}
  R_{3}^{\genuine}(p_1 ,p_2, p_3 )\equiv
                   1 + \frac{ C_{3}(p_1,p_2,p_3)}
                       {\rho_{0}(p_1,p_2,p_3)}
  \label{eqR3g} \ \ \ .
\end{equation}
The kinematical variable $Q_3$ is used to study three-particle correlations.
For a three-pion system, $Q_3=\sqrt{M_{123}^{2}-9m_{\pi}^{2}}$, with $M_{123}$ the invariant
mass of the pion triplet and $m_{\pi}$ the mass of the pion.
In this Letter, $\rho_3$ is defined as
\begin{eqnarray}
\rho_{3}(Q_3)\equiv \frac{1}{N_{\ev}}\frac{\rd n_{\triplets}}{\rd Q_3}\ \ \ ,
\end{eqnarray}
with $N_{\ev}$ the number of selected events and $n_{\triplets}$ the number of triplets of like-sign tracks,
and $\rho_2$ is defined analoguously.

Assuming totally incoherent production of particles
and a source density $f(x)$ in space-time with no dependence on the four-momentum of the emitted particle,
the BE correlation functions 
is related to the source density by~\cite{lorstad,lyub}
\begin{eqnarray}
  R_2(Q_{ij})               &=& 1+|F(Q_{ij})|^2    \label{eqr2}   \\
  R_3(Q_{12},Q_{23},Q_{31}) &=& 1+|F(Q_{12})|^2+|F(Q_{23})|^2+|F(Q_{31})|^2 \nonumber  \\
                            & & \phantom{1} +2\,\Re\{F(Q_{12})F(Q_{23})F(Q_{31})\}  \label{eq9} \\
  R_{3}^{\genuine}(Q_{12},Q_{23},Q_{31}) &=& 1+2\,\Re\{F(Q_{12})F(Q_{23})F(Q_{31})\}\label{eq10}
\end{eqnarray}
where $F(Q_{ij})$ is the Fourier transform of $f(x)$.

$R_2$ does not depend on the phase $\phi_{ij}$ contained in 
$F(Q_{ij})\equiv |F(Q_{ij})|\,\exp(\iota\phi_{ij})$.
However, this phase survives in the three-particle BE correlation functions, Equations~(\ref{eq9}) and~(\ref{eq10}). 
Assuming fully incoherent particle production,
the phase $\phi_{ij}$ can be non-zero 
only if the space-time distribution of the source is asymmetric and $Q_{ij}>0$. 
Defining
\begin{equation}
  \omega(Q_{12},Q_{23},Q_{31}) = \frac{R_{3}^{\genuine}(Q_{12},Q_{23},Q_{31})-1}
                            {2\sqrt{(R_{2}(Q_{12})-1)(R_{2}(Q_{23})-1)(R_{2}(Q_{31})-1)}} \ \ \ ,
  \label{eq11} 
\end{equation}
then for an incoherent source Equations~(\ref{eqr2}) and~(\ref{eq10}) imply that
$\omega=\cos\phi$, where $\phi\equiv\phi_{12}+\phi_{23}+\phi_{31}$.
Furthermore, as $Q_{ij}\rightarrow0$, then $\phi_{ij}\rightarrow0$, 
and hence $\omega\rightarrow1$.
For  $Q_{ij}>0$, a deviation from unity can be caused by an asymmetry in the production.
However, this will only result in a small (a few percent) reduction of
$\omega$~\cite{heiselbheiselb2,zhang}, and this only in the case where
the asymmetry occurs around the point of highest emissivity.
It is important to emphasize that for (partially) coherent sources, $\omega$ can still be defined by
Equation~(\ref{eq11}), but Equations~(\ref{eqr2}--\ref{eq10}) are no longer valid,
in which case more complicated expressions are needed~\cite{zhang},
and one can no longer deduce that
$\omega=\cos\phi$ or that $\omega\rightarrow1$ as $Q_{ij}\rightarrow0$.
In at least one type of model, one can make the stronger statement that the limit $\omega=1$
at $Q_{ij}\rightarrow0$ can only be reached if the source is fully incoherent\cite{corehalo}.

\section*{Determination of {\boldmath $R_3$} and {\boldmath $R_3^{\genuine}$}}
The reference sample, from which $\rho_0$ is determined, is formed by mixing particles from different
data events in the following way. 
Firstly, 1000 events are rotated to a system with the $z$-axis
along the thrust axis and are stored in a ``pool''. Then, tracks of each new event outside the
pool are exchanged with tracks of the same charge from events in
the pool having about the same (within about 20\%) multiplicity,
under the condition that all tracks originate from different events. 
Thus, after this procedure the new event consists of tracks originating from different events
in the pool, and its original tracks have entered the pool.
This updating process prevents any regularities in the reference sample.
Finally, $Q_3$ is calculated for each triplet of like-sign tracks, resulting in the density $\rho_{\mix}$.

This mixing procedure removes more correlations than just those of BE, \eg, those from energy-momentum
conservation and from resonances. This effect is estimated using a \MC\  model with no BE effects (JETSET
or HERWIG) at generator level and using pions only. 
Thus, in the absence of BEC, the corrected three-particle density is given by
\begin{equation}
  \rho_0 (Q_3)=\rho_{\mix}(Q_3) \mathcal{C}_{\mix}(Q_3) , \ \ \ \where \ \ \mathcal{C}_{\mix}(Q_3)=\left[
      \frac{\rho_3(Q_3)}{\rho_{\mix}(Q_3)}\right]_{\MC,\noBE} \ \ \ .
  \label{refsample}
\end{equation}

The density $\rho_3$, measured in data, must be corrected for detector resolution, acceptance,
efficiency and for particle misidentification. For this we use a multiplicative factor, $\mathcal{C}_{\det}$, derived from
\MC\  studies. Since no hadrons are identified in the analysis, $\mathcal{C}_{\det}$ is given by
the ratio of the three-pion correlation function found from \MC\  events at generator level to the three-particle
correlation function found using all particles after full detector simulation, reconstruction and 
selection.
Combining this correction factor with Equations~(\ref{threecorr}) and (\ref{refsample}) results in
\begin{equation}
   R_{3}(Q_3) = \frac{\rho_3(Q_3)     \, \mathcal{C}_{\det}(Q_3)}
                     {\rho_{\mix}(Q_3)\, \mathcal{C}_{\mix}(Q_3)}  \ \ \ . \label{eqr3_2}
\end{equation}

The genuine three-particle BE correlation function, $R_{3}^{\genuine}$, is obtained via
\begin{equation}
   R_3^{\genuine}=R_3-R_{1,2}+1 \ \ \ ,     \label{eqr3genfunc}
\end{equation}
where $R_{1,2}\equiv (\sum\rho_{1}\rho_{2})/\rho_0 -2$ is the contribution due to two-particle correlations, 
as may be seen from Equations~(\ref{eqr3}) and~(\ref{threecorr}).
The product of densities $\sum\rho_{1}(p_1)\rho_{2}(p_2,p_3)$ is determined by a similar mixing procedure,
as defined earlier, where two like-sign tracks from the same event are combined with 
one track having the same charge from another event with the same multiplicity. 
Finally, the variable $Q_3$ is calculated from these three
tracks. This procedure is similar to that given in Reference~\citen{na22}.
The ratio $(\sum\rho_{1}\rho_{2})/\rho_0$ is also corrected for 
detector effects as $\rho_3/\rho_{\mix}$.

In our analysis, we use JETSET without BEC and HERWIG to determine $\mathcal{C}_{\mix}$ and
JETSET with and without BEC as well as HERWIG to determine $\mathcal{C}_{\det}$. 
These six \MC\ combinations serve to estimate systematic uncertainties.
The corrections are largest at small $Q_3$.  
At $Q_3=0.16\,\GeV$, these corrections to $R_3$ are
$\mathcal{C}_{\mix}\approx 5$--30\% and $\mathcal{C}_{\det}\approx 20$--30\%, 
depending on which \MC\ is used.
These corrections for $R_3$ and $R_{1,2}$ are correlated and largely cancel in calculating
$R_3^{\genuine}$ by Equation~(\ref{eqr3genfunc}).

To correct the data for two-pion Coulomb repulsion in calculating $\rho_2$, each 
pair of pions is weighted by the inverse Gamow factor~\cite{gamow}
\begin{equation}
  G_2^{-1}(\eta_{ij})=\frac{\exp(2\pi\eta_{ij})-1}{2\pi\eta_{ij}}, \ \ \ \where \ \
  \eta_{ij}=\frac{m_{\pi}\alpha}{Q_{ij}} 
\end{equation}
and $\alpha$ is the fine-structure constant.
It has been shown~\cite{lors} that this Gamow factor is an approximation suitable for our purposes.
For $\rho_3$, the weight of each triplet is taken as the product of the weights of the three pairs within it.
For $\sum\rho_2\rho_1$ we use the same weight but with $G_2(Q_{ij})\equiv 1$ when 
particles $i$ and $j$ come from different events. At the lowest $Q_3$ values under consideration,
the Coulomb correction is approximately 10\%, 3\% and 2\%, for $\rho_3$, $\sum\rho_{1}\rho_{2}$ and $\rho_2$,
respectively.

\section*{Results} 
\begin{figure}
\begin{center}
\epsfig{figure=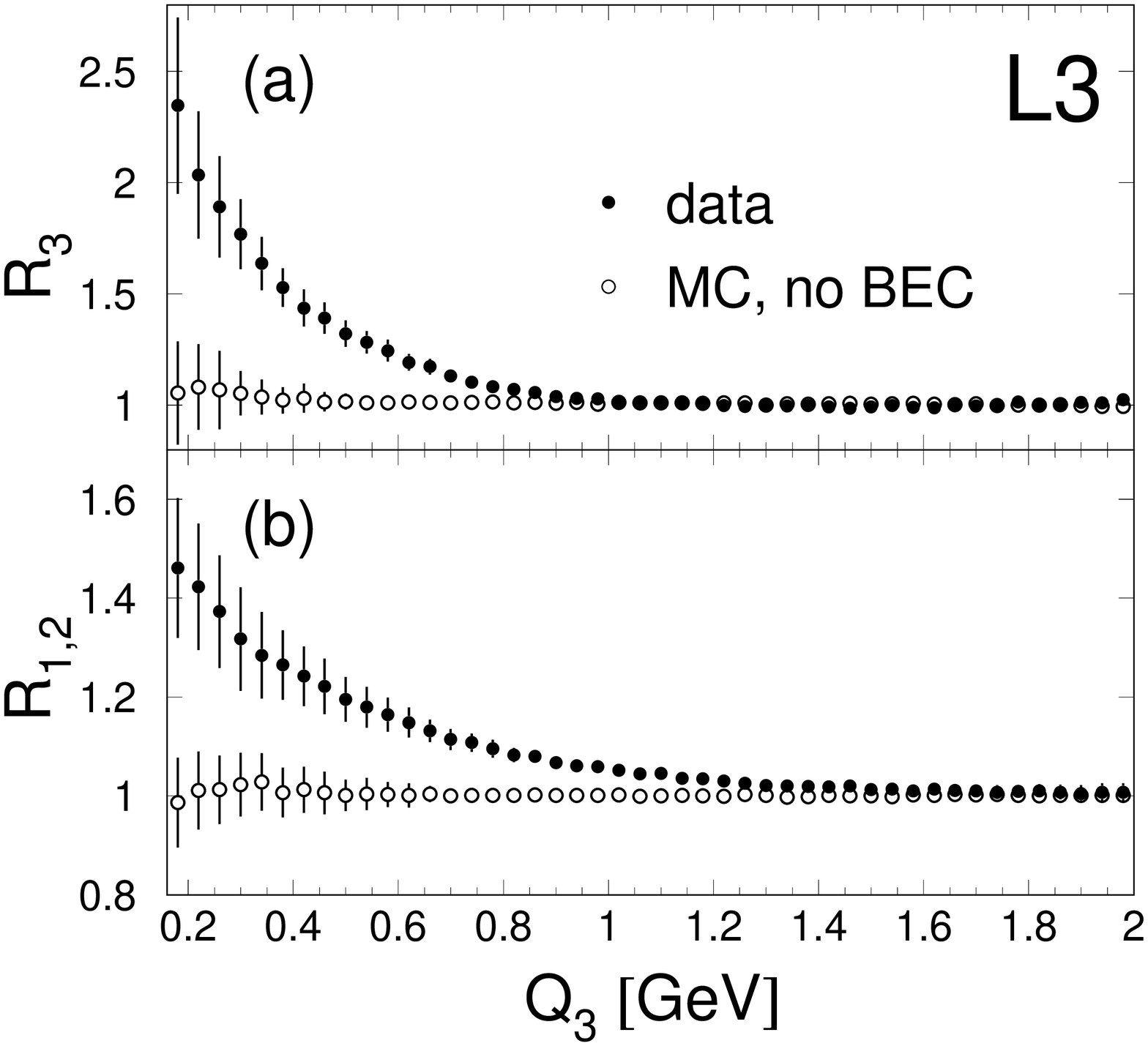, width=.864\linewidth}\\

\vspace{-1.5cm}
\epsfig{figure=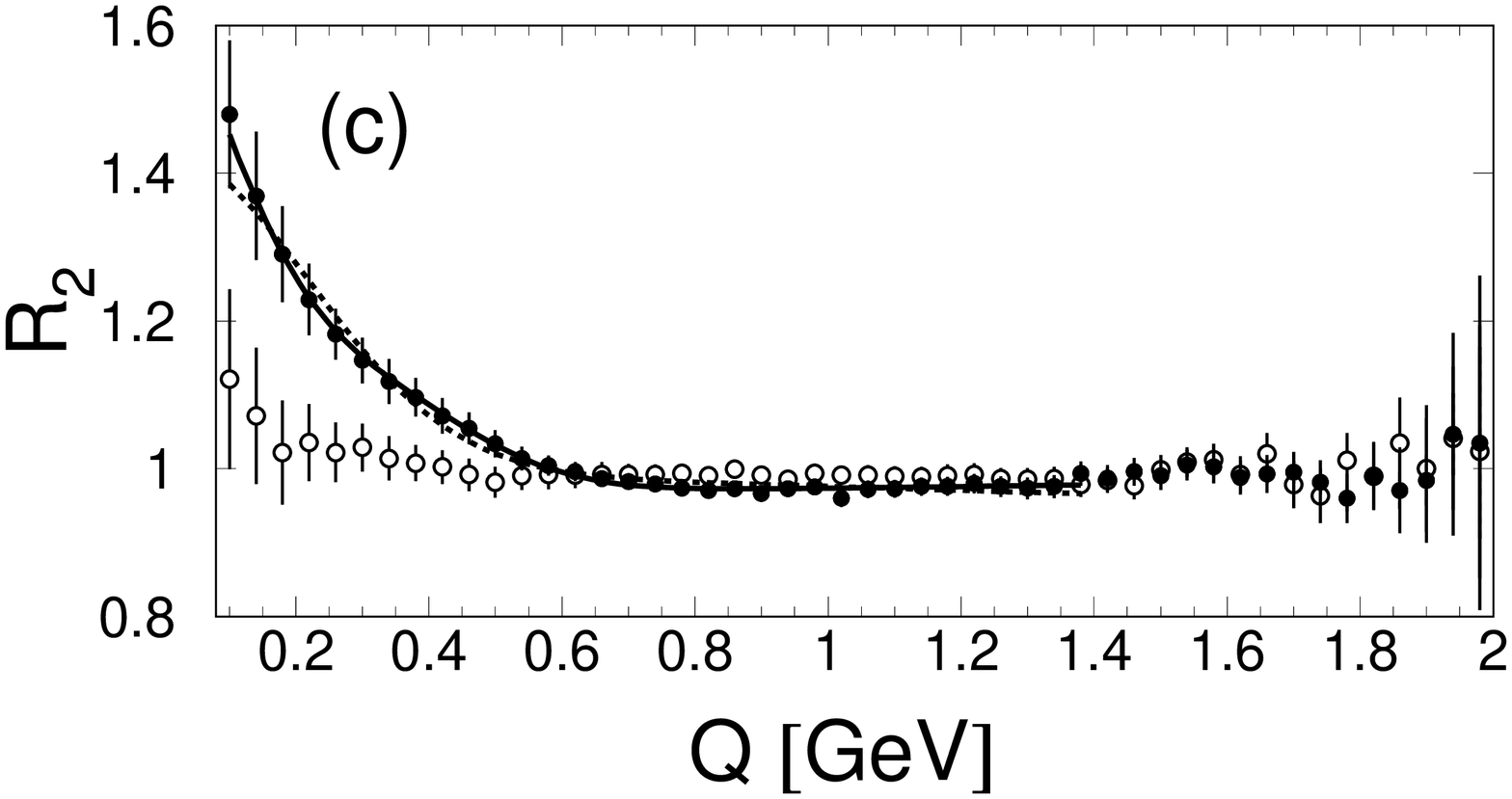, width=.864\linewidth}
\caption{(a) The three-particle BE correlation function, $R_3$, from Equation~(\ref{eqr3_2}), 
(b) the contribution of two-particle correlations, $R_{1,2}\equiv(\sum\rho_2\rho_1)/\rho_{0}-2$,
and (c) $R_2$ from Equation~(\ref{eqr2}).
The full circles correspond to the data and the error bars to $\sigma_1$ (see text).
The open circles correspond to the results from \MC\  models without BEC.
In (c) the dashed and full lines show the fits of Equations~(\ref{paramr2}) and ~(\ref{paramr2ad}),
respectively.}
\label{fig4}
\end{center}
\end{figure}

The measurements of $R_3$, $R_{1,2}$ and $R_2$ are shown in Figure~\ref{fig4}.
The full circles correspond to the averages of the data points obtained from the six possible \MC\ 
combinations used to determine $\mathcal{C}_{\mix}$ and $\mathcal{C}_{\det}$.
The error bars, $\sigma_1$, include both the statistical uncertainty and the systematic uncertainty of the
\MC\  modeling, which is taken as the r.m.s.\ of the values obtained using the different \MC\  combinations.
This dominant systematic uncertainty is, for $Q_3 <0.8\,\GeV$,
about a factor 5 to 7 larger than the statistical uncertainty and is correlated between the $R_3$, $R_{1,2}$
and $R_2$ distributions of Figure~\ref{fig4} and between bins. Figure~\ref{fig4}a shows the existence of
three-particle correlations and from Figure~\ref{fig4}b it is clear that about half is due to two-particle 
correlations.  Figure~\ref{fig4}c shows the two-particle correlations.

As a check, $R_3$, $R_{1,2}$ and $R_2$ are also computed for \MC\  models without BEC, both
HERWIG and JETSET, after detector simulation, reconstruction and selection. 
For the mixing and detector corrections
all possible \MC\  combinations, giving non-trivial results, are studied. The results of this check are shown 
in Figure~\ref{fig4} as open circles and, as expected, flat distributions around unity are observed.

Figure~\ref{fig5}a shows the genuine three-particle BE correlation function $R_3^{\genuine}$.
The data points show the existence of genuine three-particle BE correlations.
The \MC\  systematic uncertainty is highly correlated from bin to bin. At $Q_3 <0.8\,\GeV$, it is
about a factor 1.5 to 3.5 larger than the statistical uncertainty, the higher value corresponding to
the lowest $Q_3$ value used. The open circles correspond to \MC\  without BEC and form a flat distribution around
unity, as expected.

\subsection*{Gaussian Parametrizations}
A fit from $Q_3=0.16$ to 1.40\,\GeV\ using the covariance matrix including
both the statistical uncertainty and the systematic uncertainty
due to the \MC\  modeling, $\sigma_1$, is performed on the data points with the commonly 
used~\cite{lorstad,na22,delphicol,opalcol} parametrization
\begin{equation}
  R_3^{\genuine}(Q_3) = \tilde{\gamma} \left[ 1+2\tilde{\lambda}^{1.5}\exp(-\tilde{R}^2 Q_3^2/2)\right]
                        (1+\tilde{\varepsilon} Q_3) \ \ \ ,
\label{param}
\end{equation}
where $\tilde{\gamma}$ is an overall normalization factor, $\tilde{\lambda}$ measures the strength of the correlation,
$\tilde{R}$ is a measure for the effective source size in space-time and the term $(1+\tilde{\varepsilon} Q_3)$
takes into account possible long-range momentum correlations. 
The form of this parametrization is a consequence of the assumptions that $\omega=1$ and that 
$|F(Q_{ij})| = \sqrt{\lambda}\,\exp(-\tilde{R}^2 Q_{ij}^2 /2)$,
as would be expected for a Gaussian source density.
The fit results are given in the first column of Table~\ref{tab1} and shown as the full line in Figure~\ref{fig5}a.

In addition to the \MC\  modeling, we investigate four other sources of systematic uncertainties 
on the fit parameters. Firstly, the influence of a different mixing sample is studied by removing 
the conditions that tracks are replaced by tracks with the same charge and coming from events with 
approximately the same multiplicity. For each of the six \MC\  combinations, the difference in the 
fit results between the two mixing methods is taken as an estimate of the systematic uncertainty. 
The square root of the mean of the squares of these differences is assigned as the systematic 
uncertainty from this source. In the same way, systematic uncertainties related to track and event 
selection and to the choice of the fit range are evaluated. The analysis is repeated with stronger 
and weaker selection criteria, changing the number of events by about $\pm$11\% and the number of 
tracks by about $\pm$12\%. The fit range is varied by removing the first point of the fit and 
varying the end point by $\pm 200\,\MeV$. Finally, we study the influence of removing like-sign 
track pairs with small polar and azimuthal opening angles. The maximum deviation that is found by 
varying the cuts on these angles up to 6$^{\circ}$, is taken as the systematic uncertainty from 
this source. The total systematic uncertainty due to these four sources is obtained by adding the 
four uncertainties in quadrature. We refer to this systematic uncertainty as $\sigma_2$. For all fit 
parameters, the largest part of the total uncertainty is due to the six possible combinations of mixing and 
detector \MC\  corrections and amounts to 50 to 90\%. Table~\ref{tab3} shows the uncertainties for each of the 
sources for the fit parameters of Equation~(\ref{param}). 

As a cross-check, the analysis is repeated without the momentum cut of 1\,\GeV\ and without the cut of
3$^{\circ}$
on the opening angle of like-sign track pairs. The results agree with those given in Table~\ref{tab1}
well within quoted uncertainties, but the systematic uncertainties are approximately twice as large.

To measure the ratio $\omega$, we also need to determine the two-particle 
BE correlation function $R_2 (Q)$.
This is done in the same way as the three-particle BE correlation function. 
The correlation function $R_2$ is parametrized as a Gaussian:
\begin{equation}
  R_2 (Q)=\gamma \left[ 1+\lambda\exp(-R^2 Q^2)\right] (1+\varepsilon Q) \ \ \ .
\label{paramr2}
\end{equation}
The parametrization starts at $Q=0.08\,\GeV$, consistent with the study of $R_3$ from $Q_{3}=0.16\,\GeV$.
The fit results\footnote{Due to the use of a different fit range, these fit results differ from those found in
                         Reference~\citen{michiel}. The same fit range gives similar results.}
are given in the first column of Table~\ref{tab2} and in Figure~\ref{fig4}c.

If the space-time structure of the pion source is Gaussian
and the pion production mechanism is completely incoherent, $\tilde{\lambda}$ and $\tilde{R}$ as
derived from the fit to Equation~(\ref{param}) measure the same correlation strength 
and effective source size as $\lambda$ and $R$ of Equation~(\ref{paramr2}).
The values of $\lambda$ and $R$ are consistent with $\tilde{\lambda}$ and $\tilde{R}$,
as expected for fully incoherent production of pions ($\omega=1$).
Using the values of $\lambda$ and $R$ instead of $\tilde{\lambda}$ and $\tilde{R}$ in Equation~(\ref{param}),
which is justified if $\omega=1$, results in the dashed line in Figure~\ref{fig5}a. It is only slightly different
from the result of the fit to Equation~(\ref{param}), indicating that $\omega$ is indeed near unity.

Another way to see how well $R_3^{\genuine}$ corresponds to a completely incoherent pion production interpretation 
and a Gaussian source density in space-time, is to compute
$\omega$ with Equation~(\ref{eq11}), for each bin in $Q_3$ (from 0.16 to 0.80\,\GeV), using the
measured $R_3^{\genuine}$ and $R_2$ derived from the parametrization of Equation~(\ref{paramr2}). The result is shown
in Figure~\ref{fig11}a. At low $Q_3$, $\omega$ appears to be higher than unity.

\subsection*{Extended Gaussian Parametrizations}
However, the assumption of a Gaussian source density is only an approximation, as observed in
Reference~\citen{elong} and confirmed by the $\chi^2$ of the fit to Equation~(\ref{paramr2}). 
Deviations from a Gaussian can be studied by expanding in terms of derivatives of the Gaussian, 
which are related to Hermite polynomials. Taking only the lowest-order non-Gaussian term into account, 
this so-called Edgeworth expansion~\cite{edge1edge2} replaces
the parametrization of Equation~(\ref{paramr2}) by 
\begin{equation}
  R_2 (Q) = \gamma \left[ 1+\lambda\exp(-R^2 Q^2)(1+\kappa H_3 (\sqrt{2}RQ)/6)\right] (1+\varepsilon Q) \ \ \ ,
  \label{paramr2ad}
\end{equation}
where $\kappa$ measures the deviation from the Gaussian and $H_3 (x)\equiv x^3
-3x$ is the third-order Hermite polynomial.
The fit results for the two-particle BE correlation function with this parametrization
are given in the second column of Table~\ref{tab2}. 

Using the first-order Edgeworth expansion of the Gaussian, Equation~(\ref{paramr2ad}), and using Equation~(\ref{eq11}),
assuming $\omega=1$, the parametrization of Equation~(\ref{param}) becomes
\begin{eqnarray}
  R_3^{\genuine}(Q_3)\!\!\!&=&\!\!\!\tilde{\gamma} \left(1+2\tilde{\lambda}^{1.5}\exp(-\tilde{R}^2 Q_3^2/2)
     \left[\prod_{i,j=1,\,j>i}^{3}\sqrt{ 1
     +\frac{H_3 (\sqrt{2}\tilde{R} Q_{ij})}{6}\tilde{\kappa}\,}\,\,\right] \right)\left(
    1+\tilde{\varepsilon} Q_3\right) \nonumber \\
  &\simeq& \!\!\!\tilde{\gamma} \left(1+2\tilde{\lambda}^{1.5}\exp(-\tilde{R}^2 Q_3^2/2)\left[1
    +\frac{H_3 (\sqrt{2}\tilde{R} Q_3 /2)}{6}\tilde{\kappa}\right]^{1.5} \right)\left(1 +\tilde{\varepsilon} Q_3\right)
  \ \ \ . \label{paramad}
\end{eqnarray}
In the second line the approximation is made that $Q_{ij}=Q_3 /2$.
The effect of this approximation on $R_3^{\genuine}$ is small compared to the statistical uncertainty.
The results of a fit to Equation~(\ref{paramad}) are given in the
second column of Table~\ref{tab1}. The uncertainties are summarized in Table~\ref{tab3}.

\begin{figure}
\begin{center}
\epsfig{figure=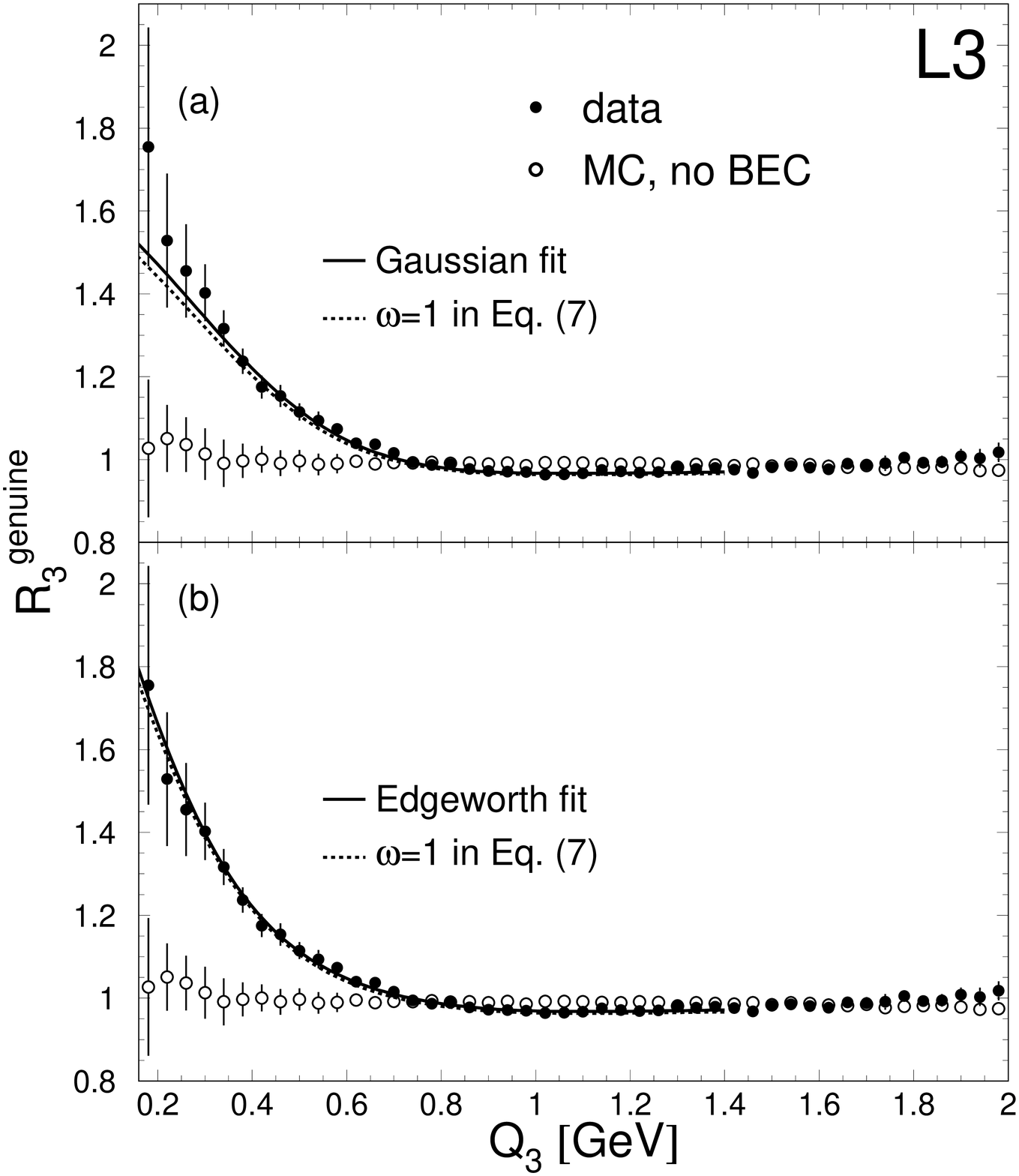, width=.945\linewidth}
\caption{The genuine three-particle BE correlation function $R_3^{\mathrm{genuine}}$, Equation~(\ref{eqr3genfunc}).
The full circles correspond to the data and the error bars to $\sigma_1$.
The open circles correspond to results from \MC\ models without BEC.
In (a) the full line shows the fit of Equation~(\ref{param}), the dashed line the prediction of completely
incoherent pion production and a Gaussian source density in space-time, derived from parametrizing
$R_2$ with Equation~(\ref{paramr2}). In (b) Equations~(\ref{paramad}) and~(\ref{paramr2ad}) are used, respectively.}
\label{fig5}
\end{center}
\end{figure}

\begin{figure}
\begin{center}
\epsfig{figure=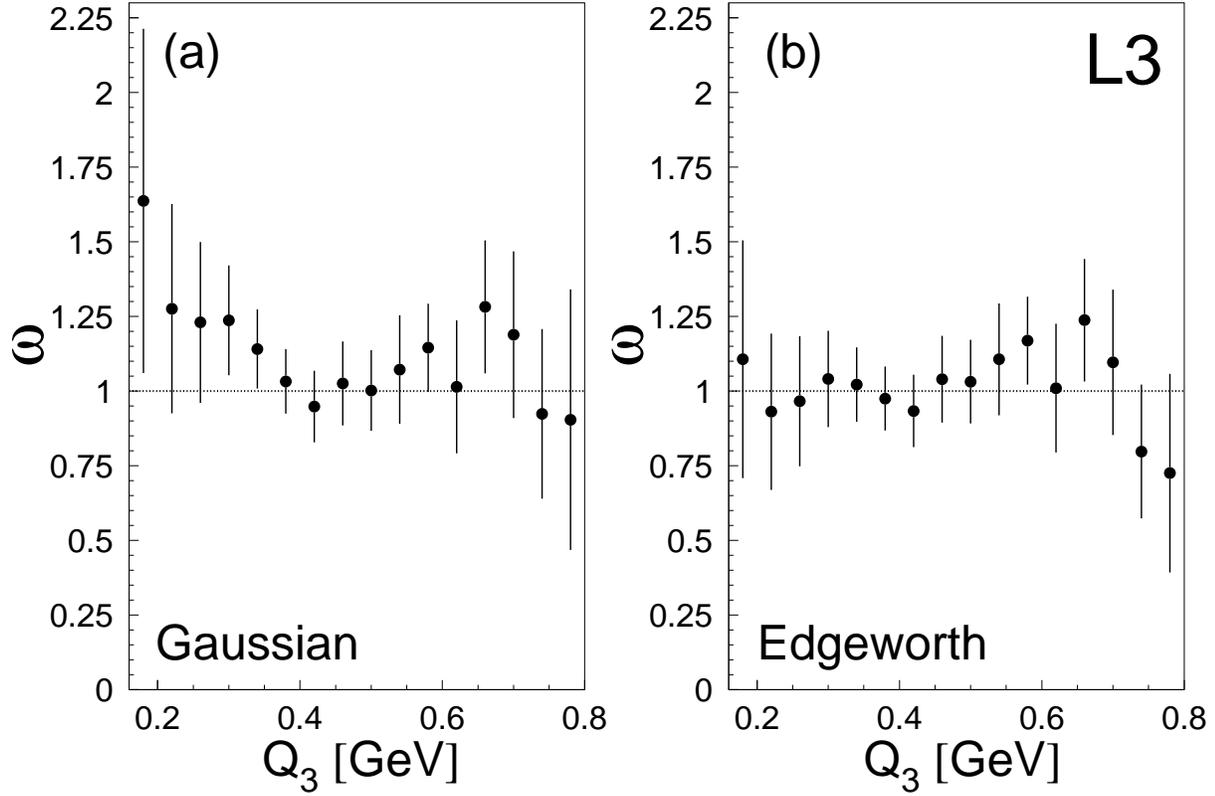,width=1.0\linewidth}
\caption{The ratio $\omega$ as a function of $Q_3$ assuming $R_2$ is described (a) by the Gaussian, Equation~(\ref{paramr2}),
and (b) by the first-order Edgeworth expansion of the Gaussian, Equation~(\ref{paramr2ad}).
The error bars correspond to $\sigma_1$.
For completely incoherent production, $\omega=1$.
}
\label{fig11}
\end{center}
\end{figure}

For both $R_3^{\genuine}$ and $R_2$, a better $\chi^2$/NDF is found using the Edgeworth expansion,
and the values of $\tilde{\lambda}$ and $\lambda$ are significantly higher, as shown in
Tables~\ref{tab1} and~\ref{tab2} and in Figures~\ref{fig5}b and~\ref{fig4}c.
The values for $\tilde{\lambda}$ and $\tilde{R}$ are still consistent with the corresponding $\lambda$ and
$R$, as would be expected for a fully incoherent production mechanism of pions.

In Figure~\ref{fig5}b, as in Figure~\ref{fig5}a, we observe good agreement
between the fit of $R_3^{\genuine}$ using the parametrization of Equation~(\ref{paramad}) and the prediction of a completely
incoherent pion production mechanism, derived from parametrizing $R_2$ with Equation~(\ref{paramr2ad}), over the
full range of $Q_3$. In Figure~\ref{fig11}b, no deviation from unity is observed for the ratio $\omega$.
This indicates that the data agree with the assumption of fully incoherent pion production.

Fits to samples generated with JETSET with BE effects modelled by $\mathrm{BE}_0$ or
$\mathrm{BE}_{32}$\footnote{The BE$_{32}$ algorithm uses the values PARJ(92)=1.68 and PARJ(93)=0.38\,\GeV.}~\cite{jetset32} 
result in values of $\tilde{R}$ in agreement with the data 
but in significantly higher values of $\tilde{\lambda}$.
This confirms the observation in Figure~\ref{fig1b}f that
the standard BE implementations of JETSET overestimate the genuine three-particle BEC.

\section*{Summary}
Three-particle, as well as two-particle Bose-Einstein correlations of like-sign charged pions
have been measured in
hadronic Z decay.  Genuine three-particle BE correlations are observed.  The correlation
functions are better parametrized by an Edgeworth expansion of a Gaussian than by a
simple Gaussian.  Combining the two- and three-particle correlations shows that the data
are consistent with a fully incoherent production mechanism of pions.

\section*{Acknowledgements}
Clarifying discussions with T. Cs\"org\H{o} are gratefully acknowledged.

\newpage
\typeout{   }     
\typeout{Using author list for paper 256 -  }
\typeout{$Modified: Jul 15 2001 by smele $}
\typeout{!!!!  This should only be used with document option a4p!!!!}
\typeout{   }
%
%
%
%
%
%

\newcount\tutecount  \tutecount=0
\def\tutenum#1{\global\advance\tutecount by 1 \xdef#1{\the\tutecount}}
\def\tute#1{$^{#1}$}
\tutenum\aachen            
\tutenum\nikhef            
\tutenum\mich              
\tutenum\lapp              
\tutenum\basel             
\tutenum\lsu               
\tutenum\beijing           
\tutenum\berlin            
\tutenum\bologna           
\tutenum\tata              
\tutenum\ne                
\tutenum\bucharest         
\tutenum\budapest          
\tutenum\mit               
\tutenum\panjab            
\tutenum\debrecen          
\tutenum\dublin            
\tutenum\florence          
\tutenum\cern              
\tutenum\wl                
\tutenum\geneva            
\tutenum\hefei             
\tutenum\lausanne          
\tutenum\lyon              
\tutenum\madrid            
\tutenum\florida           
\tutenum\milan             
\tutenum\moscow            
\tutenum\naples            
\tutenum\cyprus            
\tutenum\nymegen           
\tutenum\caltech           
\tutenum\perugia           
\tutenum\peters            
\tutenum\cmu               
\tutenum\potenza           
\tutenum\prince            
\tutenum\riverside         
\tutenum\rome              
\tutenum\salerno           
\tutenum\ucsd              
\tutenum\sofia             
\tutenum\korea             
\tutenum\purdue            
\tutenum\psinst            
\tutenum\zeuthen           
\tutenum\eth               
\tutenum\hamburg           
\tutenum\taiwan            
\tutenum\tsinghua          

{
\parskip=0pt
\noindent
{\bf The L3 Collaboration:}
\ifx\selectfont\undefined
 \baselineskip=10.8pt
 \baselineskip\baselinestretch\baselineskip
 \normalbaselineskip\baselineskip
 \ixpt
\else
 \fontsize{9}{10.8pt}\selectfont
\fi
\medskip
\tolerance=10000
\hbadness=5000
\raggedright
\hsize=162truemm\hoffset=0mm
\def\r{\rlap,}
\noindent

P.Achard\r\tute\geneva\ 
O.Adriani\r\tute{\florence}\ 
M.Aguilar-Benitez\r\tute\madrid\ 
J.Alcaraz\r\tute{\madrid,\cern}\ 
G.Alemanni\r\tute\lausanne\
J.Allaby\r\tute\cern\
A.Aloisio\r\tute\naples\ 
M.G.Alviggi\r\tute\naples\
H.Anderhub\r\tute\eth\ 
V.P.Andreev\r\tute{\lsu,\peters}\
% ...
\newpage
\begin{list}{A}{\itemsep=0pt plus 0pt minus 0pt\parsep=0pt plus 0pt minus 0pt
                \topsep=0pt plus 0pt minus 0pt}
\item[\aachen]
 I. Physikalisches Institut, RWTH, D-52056 Aachen, FRG$^{\S}$\\
 III. Physikalisches Institut, RWTH, D-52056 Aachen, FRG$^{\S}$
\item[\nikhef] National Institute for High Energy Physics, NIKHEF, 
     and University of Amsterdam, NL-1009 DB Amsterdam, The Netherlands
\item[\mich] University of Michigan, Ann Arbor, MI 48109, USA
\item[\lapp] Laboratoire d'Annecy-le-Vieux de Physique des Particules, 
     LAPP,IN2P3-CNRS, BP 110, F-74941 Annecy-le-Vieux CEDEX, France
\item[\basel] Institute of Physics, University of Basel, CH-4056 Basel,
     Switzerland
\item[\lsu] Louisiana State University, Baton Rouge, LA 70803, USA
\item[\beijing] Institute of High Energy Physics, IHEP, 
  100039 Beijing, China$^{\triangle}$ 
\item[\berlin] Humboldt University, D-10099 Berlin, FRG$^{\S}$
\item[\bologna] University of Bologna and INFN-Sezione di Bologna, 
     I-40126 Bologna, Italy
\item[\tata] Tata Institute of Fundamental Research, Mumbai (Bombay) 400 005, India
\item[\ne] Northeastern University, Boston, MA 02115, USA
\item[\bucharest] Institute of Atomic Physics and University of Bucharest,
     R-76900 Bucharest, Romania
\item[\budapest] Central Research Institute for Physics of the 
     Hungarian Academy of Sciences, H-1525 Budapest 114, Hungary$^{\ddag}$
\item[\mit] Massachusetts Institute of Technology, Cambridge, MA 02139, USA
\item[\panjab] Panjab University, Chandigarh 160 014, India.
\item[\debrecen] KLTE-ATOMKI, H-4010 Debrecen, Hungary$^\P$
\item[\dublin] Department of Experimental Physics,
  University College Dublin, Belfield, Dublin 4, Ireland
\item[\florence] INFN Sezione di Firenze and University of Florence, 
     I-50125 Florence, Italy
\item[\cern] European Laboratory for Particle Physics, CERN, 
     CH-1211 Geneva 23, Switzerland
\item[\wl] World Laboratory, FBLJA  Project, CH-1211 Geneva 23, Switzerland
\item[\geneva] University of Geneva, CH-1211 Geneva 4, Switzerland
\item[\hefei] Chinese University of Science and Technology, USTC,
      Hefei, Anhui 230 029, China$^{\triangle}$
\item[\lausanne] University of Lausanne, CH-1015 Lausanne, Switzerland
\item[\lyon] Institut de Physique Nucl\'eaire de Lyon, 
     IN2P3-CNRS,Universit\'e Claude Bernard, 
     F-69622 Villeurbanne, France
\item[\madrid] Centro de Investigaciones Energ{\'e}ticas, 
     Medioambientales y Tecnol\'ogicas, CIEMAT, E-28040 Madrid,
     Spain${\flat}$ 
\item[\florida] Florida Institute of Technology, Melbourne, FL 32901, USA
\item[\milan] INFN-Sezione di Milano, I-20133 Milan, Italy
\item[\moscow] Institute of Theoretical and Experimental Physics, ITEP, 
     Moscow, Russia
\item[\naples] INFN-Sezione di Napoli and University of Naples, 
     I-80125 Naples, Italy
\item[\cyprus] Department of Physics, University of Cyprus,
     Nicosia, Cyprus
\item[\nymegen] University of Nijmegen and NIKHEF, 
     NL-6525 ED Nijmegen, The Netherlands
\item[\caltech] California Institute of Technology, Pasadena, CA 91125, USA
\item[\perugia] INFN-Sezione di Perugia and Universit\`a Degli 
     Studi di Perugia, I-06100 Perugia, Italy   
\item[\peters] Nuclear Physics Institute, St. Petersburg, Russia
\item[\cmu] Carnegie Mellon University, Pittsburgh, PA 15213, USA
\item[\potenza] INFN-Sezione di Napoli and University of Potenza, 
     I-85100 Potenza, Italy
\item[\prince] Princeton University, Princeton, NJ 08544, USA
\item[\riverside] University of Californa, Riverside, CA 92521, USA
\item[\rome] INFN-Sezione di Roma and University of Rome, ``La Sapienza",
     I-00185 Rome, Italy
\item[\salerno] University and INFN, Salerno, I-84100 Salerno, Italy
\item[\ucsd] University of California, San Diego, CA 92093, USA
\item[\sofia] Bulgarian Academy of Sciences, Central Lab.~of 
     Mechatronics and Instrumentation, BU-1113 Sofia, Bulgaria
\item[\korea]  The Center for High Energy Physics, 
     Kyungpook National University, 702-701 Taegu, Republic of Korea
\item[\purdue] Purdue University, West Lafayette, IN 47907, USA
\item[\psinst] Paul Scherrer Institut, PSI, CH-5232 Villigen, Switzerland
\item[\zeuthen] DESY, D-15738 Zeuthen, 
     FRG
\item[\eth] Eidgen\"ossische Technische Hochschule, ETH Z\"urich,
     CH-8093 Z\"urich, Switzerland
\item[\hamburg] University of Hamburg, D-22761 Hamburg, FRG
\item[\taiwan] National Central University, Chung-Li, Taiwan, China
\item[\tsinghua] Department of Physics, National Tsing Hua University,
      Taiwan, China
\item[\S]  Supported by the German Bundesministerium 
        f\"ur Bildung, Wissenschaft, Forschung und Technologie
\item[\ddag] Supported by the Hungarian OTKA fund under contract
numbers T019181, F023259 and T037350.
\item[\P] Also supported by the Hungarian OTKA fund under contract
  number T026178.
\item[$\flat$] Supported also by the Comisi\'on Interministerial de Ciencia y 
        Tecnolog{\'\i}a.
\item[$\sharp$] Also supported by CONICET and Universidad Nacional de La Plata,
        CC 67, 1900 La Plata, Argentina.
\item[$\triangle$] Supported by the National Natural Science
  Foundation of China.
\end{list}
}
\vfill



\newpage
\begin{table}
\centering
\begin{tabular}{|c|c|c|}
\hline
 parameter & Eq.~(\ref{param}) & Eq.~(\ref{paramad}) \\
\hline
$\tilde{\gamma}$    & $0.96\pm0.03\pm0.02$ & $0.95\pm0.03\pm0.02$ \\
$\tilde{\lambda}$   & $0.47\pm0.07\pm0.03$ & $0.75\pm0.10\pm0.03$ \\
$\tilde{R}$, fm     & $0.65\pm0.06\pm0.03$ & $0.72\pm0.08\pm0.03$ \\
$\tilde{\varepsilon}$, $\GeV^{-1}$ & $0.02\pm0.02\pm0.02$ & $0.02\pm0.02\pm0.02$ \\
$\tilde{\kappa}$    &        -               & $0.79\pm0.26\pm0.15$ \\
$\chi^2$/NDF& 29.9/27                & 17.7/26              \\
\hline
\end{tabular}
\caption{Values of the fit parameters for the genuine three-particle BE correlation
  function $R_3^{\mathrm{genuine}}$, using the parametrizations of Equations~(\ref{param})
  and~(\ref{paramad}). 
  The first uncertainty corresponds to $\sigma_1$, the second to $\sigma_2$, 
  defined in the text.
\label{tab1}
}
\end{table}

\begin{table}
\centering
\begin{tabular}{|c|c|c|c|c|c|c|c|c|c|c|}
\hline
 parametrization & \multicolumn{4}{|c|}{Eq.~(\ref{param})} & \multicolumn{5}{|c|}{Eq.~(\ref{paramad})} \\
\hline
 & & & & & & & & & \\[-11pt]
 fit parameter        & $\tilde{\gamma}$ & $\tilde{\lambda}$ & $\tilde{R}$, fm & $\!\tilde{\varepsilon}$, 
  $\mathrm{Ge\hspace{-.6mm}V}^{-1}\!\!$ & $\tilde{\gamma}$ & $\tilde{\lambda}$ & $\tilde{R}$, fm & $\!\tilde{\varepsilon}$, $\mathrm{Ge\hspace{-.6mm}V}^{-1}\!\!$ & $\tilde{\kappa}$ \\
\hline
\hline
 $\!\!\sigma_1 \!$ (stat.+modeling)$\!\!$ & 0.029 & 0.071 & 0.056 & 0.022 & 0.031 & 0.103 & 0.078 & 0.024 & 0.26 \\
\hline
 mixing                           & 0.004 & 0.006 & 0.009 & 0.007 & 0.010 & 0.009 & 0.011 & 0.011 & 0.04 \\
 fit range                        & 0.008 & 0.019 & 0.020 & 0.013 & 0.010 & 0.022 & 0.017 & 0.019 & 0.14 \\
 $\!$track/event sel.$\!\!$         & 0.010 & 0.013 & 0.012 & 0.008 & 0.011 & 0.016 & 0.012 & 0.007 & 0.10 \\
 $\delta\phi+\delta\theta$ cut    & 0.013 & 0.014 & 0.012 & 0.009 & 0.014 & 0.017 & 0.010 & 0.008 & 0.11 \\
\hline
 $\sigma_2$        & 0.019 & 0.028 & 0.028 & 0.020 & 0.023 & 0.033 & 0.026 & 0.024 & 0.15 \\
\hline
\end{tabular}
\caption{Contribution to the uncertainty on the fit parameters of the 
  parametrizations of Equations~(\ref{param}) and~(\ref{paramad}), respectively.
  The first uncertainty corresponds to $\sigma_1$, the others added in
  quadrature give $\sigma_2$.
  \label{tab3}
}
\end{table}

\begin{table}
\centering
\begin{tabular}{|c|c|c|}
\hline
 parameter & Eq.~(\ref{paramr2}) & Eq.~(\ref{paramr2ad}) \\
\hline
$\gamma$    & $0.98\pm0.03\pm0.02$ & $0.96\pm0.03\pm0.02$ \\
$\lambda$   & $0.45\pm0.06\pm0.03$ & $0.72\pm0.08\pm0.03$ \\
$R$, fm     & $0.65\pm0.03\pm0.03$ & $0.74\pm0.06\pm0.02$ \\
$\varepsilon$, $\GeV^{-1}$ & $0.01\pm0.01\pm0.02$ & $0.01\pm0.02\pm0.02$ \\
$\kappa$    &        -               & $0.74\pm0.21\pm0.15$ \\
$\chi^2$/NDF& 60.2/29                  & 26.0/28              \\
\hline
\end{tabular}
\caption{Values of the fit parameters for the two-particle BE correlation function, $R_2$, 
  using the parametrizations of Equations~(\ref{paramr2}) and~(\ref{paramr2ad}).
  The first uncertainty corresponds to $\sigma_1$, the second to $\sigma_2$.
  \label{tab2}
}
\end{table}


\end{document}